\documentclass{llncs}

\usepackage{graphicx} % figures

\usepackage{amsmath} % math
\usepackage{amsfonts}
\usepackage{amssymb}
\usepackage{accents}
\newcommand{\ubar}[1]{\underaccent{\bar}{#1}}

\usepackage[pdftex,dvipsnames]{xcolor}  % Coloured text etc.

\usepackage{listings} % code listings
\makeatletter
\newcommand{\printfnsymbol}[1]{%
  \textsuperscript{\@fnsymbol{#1}}%
}
\makeatother
\begin{document}

\title{Explainable nonlinear modelling of multiple time series with invertible neural networks\thanks{
The work in this paper was supported by the SFI Offshore Mechatronics
grant 237896/O30.
}
}
\titlerunning{Explainable VAR modeling with invertible NNs}

\author{Luis Miguel Lopez-Ramos\thanks{Equal contribution in terms of working hours.}%\inst{1,2}
\orcidID{0000-0001-8072-3994} \and
\\ 
Kevin Roy\printfnsymbol{2}%\inst{1,2}
%\orcidID{1111-2222-3333-4444} 
\and
Baltasar Beferull-Lozano%\inst{1,2}
\orcidID{0000-0002-0902-6245}
}

\authorrunning{Lopez-Ramos & Roy et al.}

\institute{SFI Offshore Mechatronics Center, University of Agder
\and
Intelligent Signal Processing and Wireless Networks (WISENET) Center
\and
Department of ICT, University of Agder, Grimstad, Norway
}
\maketitle

\begin{abstract}
    A method for nonlinear topology identification is proposed, based on the assumption that a collection of time series are generated in two steps: i) a vector autoregressive process in a latent space, and ii) a nonlinear, component-wise, monotonically increasing observation mapping. The latter mappings are assumed invertible, and are modeled as shallow neural networks, so that their inverse can be numerically evaluated, and their parameters can be learned using a technique inspired in deep learning. Due to the function inversion, the backpropagation step is not straightforward, and this paper explains the steps needed to calculate the gradients applying implicit differentiation. Whereas the model explainability is the same as that for linear VAR processes, preliminary numerical tests show that the prediction error becomes smaller.
\end{abstract}


\section{Introduction}
Multi-dimensional time series data are observed in many real-world systems, where some of the time series are influenced by other time series. The interrelations among the time series can be encoded in a graph structure, and identifying such structure or topology is of great interest in multiple applications \cite{giannakis2018topology}.  The inferred topology can provide insights about the underlying system and can assist in inference tasks such as prediction and anomaly detection. 

In real-world applications such as neuroscience and genomics, signal interrelations are often inherently nonlinear \cite{chen2018dynamic,fujita2010granger,shen2019nonlinear}. In these cases, using linear models may lead to inconsistent estimation of causal interactions \cite{tank2017interpretable}. We propose deep learning based methods by applying feed-forward invertible neural networks. This project proposes a low-complexity nonlinear topology identification method that is competitive with other nonlinear methods explaining time series data from a heterogeneous set of sensors.

\subsection{State of the art and contribution}

The use of linear VAR models for topology identification have been well-studied. A comprehensive review of topology identification algorithms was recently published \cite{giannakis2018topology}, where the issue of nonlinearity is discussed together with other challenges such as dynamics (meaning estimating time-varying models). 

In \cite{zaman2020online}, an efficient algorithm to estimate linear VAR coefficients from streaming data is proposed. Although the linear VAR model is not expressive enough for certain applications, it allows clear performance analysis, and is subject to continuous technical developments, such as a novel criterion for automatic order selection \cite{nassif2021automatic}, VAR estimation considering distributions different to the Gaussian, such as Student's $t$ \cite{zhou2021parameter}, or strategies to deal with missing data \cite{zhou2021parameter,ioannidis2019semiblind,zaman2020online}.

Regarding non-linear topology identification based on the VAR model, kernels are used in  \cite{shen2018online,money2021online} to linearize the nonlinear dependencies by mapping variables to a higher-dimensional Hilbert space. The growth of computational complexity and memory requirements (a.k.a.  “curse  of dimensionality”) associated with kernel representations is circumvented in \cite{shen2018online,money2021online} by  restricting  the  numeric  calculation  to  a  limited number  of  time-series  samples  using  a  time  window,  which results in suboptimal performance. A semiparametric model is proposed for the same task in \cite{farnoosh2017semiparametric}.

A different class of nonlinear topology identification methods are based on deep feedforward or recurrent NNs \cite{tank2017interpretable,tank2021neural} combined with sparsity-inducing penalties on the weights at one layer, labeled as "Granger-causality layer".

Recent work \cite{morioka2021independent} considers a nonlinear VAR framework where the innovations are not necessarily additive, and proposes estimation algorithms and identifiability results based on the assumption that the innovations are independent.

%In \cite{nassif2021automatic}, autoregressive modeling applied to electroencephalography (EEG) sleep-stage classification, and a novel criterion for automatic order selection is demonstrated.

%In \cite{zhou2021parameter}, a ML algorithm is proposed for estimation of VAR models with non-Gaussian distribution (specifically Student's $t$) with the additional challenge of missing data.

% Cite Rohan's online nonlinear topology ID

All the aforementioned nonlinear modeling techniques are based on estimating nonlinear functions that predict the future time series values in the measurement space, which entails high complexity and is not amenable to predicting multiple time instants ahead. The main contribution of our work is a modeling assumption that accounts for mild nonlinear relations that are independent of the (linear) multivariate structure, and reduces the complexity associated with long-term predictions, as explained in detail in Sec. \ref{sec:modelling}.

\section{Background}
 \subsection{Graph Topology Identification}
 
Estimating topology of a system means finding the dependencies between network data time series. These dependencies may not be physically observable; rather, there can be logical connections between data nodes that are not physically connected, but which may be (indirectly) logically connected due to, e.g. control loops. %in the overall dynamic system being monitored. 
Topology inference has the potential to contribute to the algorithmic foundations to solve important problems in signal processing (e.g. prediction, data completion, etc..) and data-driven control.

\begin{figure}[h]
\centering
\includegraphics[width=0.6\textwidth]{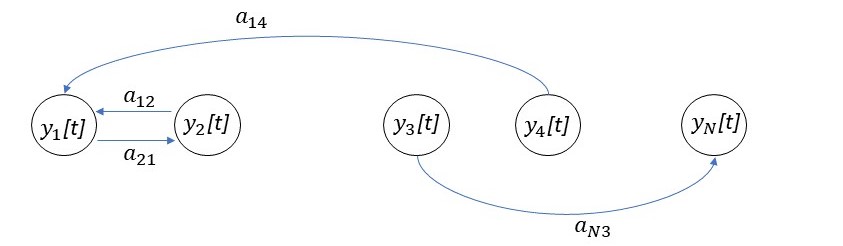}
\caption{{Illustration of an N-node network with directed edges% (in blue) 
}} 
\label{fig:node_edges}
\end{figure}

While the simplest techniques such as correlation graphs \cite{jin2020correlated} cannot determine the direction of interactions, one may also employ to this end structural equation models (SEM) or Bayesian networks \cite{yanuar2014estimation}. However, such methods account only for memory-less interactions. On the other hand, causality in the Granger \cite{granger1988concept} sense is based on the idea that the cause precedes the effect in time, and knowledge about the cause helps predicting the effect more accurately. The way Granger causality is defined makes it interesting, from a conceptual point of view, for understanding data dependencies; however, it is often computationally intractable. Thus, alternative causality definitions, such as those based on vector autoregressive (VAR) models \cite{granger1988concept,zaman2020online} are typically preferred in practical scenarios. The simplest possible VAR model is a linear VAR model. 

Consider a collection of $N$ sensors, where $y_n[t]$ denotes the measurement of the $n$-th sensor at time $t$. A $P$-th order linear VAR model can be formulated as 
\begin{align} \label{eq:var_matrix}
      y[t]=\sum_{p=1}^{P} A_p y[t-p]+u[t],  \quad \quad P \leq t \leq T
\end{align}
where $y[t]=[y_1[t],........,y_N[t]]^T$, $A_p\ \in\ R^{N\times N}$, p = 1,\ldots, P, are the matrices of VAR parameters (see Fig. 2) , $T$ is observation time period, and $u[t]={[u}_1[t],.........,u_N[t]]$ is an innovation process typically modeled as a Gaussian, temporally white random process. With $a_{n,n^\prime}^{(p)}$ being the $(n,n^\prime)$ entry of the matrix $A_p$, the r.h.s above takes the form:
% IT MAY BE A GOOD IDEA LATER TO REDUCE A BIT THE SPACING
 \begin{align} \label{eq:var_scalar}
       y_n[t] = \sum_{n^\prime=1}^{N}\sum_{p=1}^{P}{a_{n,n^\prime}^{(p)}y_{n^\prime}}[t-p]+\ u_n[t], \quad \quad P \leq t \leq T
	\end{align}
for $n = 1,\ldots, N$, %where   $a_{n,n^\prime}  = [a_{n,n^\prime}^{(1)},....,\ a_{n,n^\prime}^{(p)}]^{T}$ is the impulse response from node $n^\prime$ to node $n$; this will be a zero vector when there is no edge from node $n^\prime$ to node $n$. 
The problem of identifying a linear VAR causality model reduces to estimating the VAR coefficient matrices $\{A_p\}_{p=1}^P$ given the observations $\{y[t]\}_{t=0}^{T-1}$. The VAR causality \cite{lutkepohl2005} is determined from the support of the VAR matrix parameters and can be interpreted as a surrogate (yet not strictly equivalent) for Granger causality\footnote{Notice that VAR models encode lagged interactions, and other linear models such as structural equation models (SEM) or structural VAR (SVAR) are available if interactions at a small time scale are required. In this paper, for the sake of simplicity, we focus on learning non-linear VAR models. However, our algorithm designs can also accomodate the SEM and SVAR frameworks without much difficulty.}.
\iffalse
%\vspace{0px}
\begin{figure}[h]
\centering
\includegraphics[width=0.7\textwidth]{figures/figvar.png}
\caption{{ Tensor $A$ collecting the VAR parameter matrices \cite{zaman2020online}.  }}
\label{fig:A_tensor}
\end{figure}
\fi
%Notice that VAR models rely on uniformly-sampled data, and sometimes, one is interested in finding interactions that occur at a time scale that is smaller than the sampling rate; we refer to these interactions as “instantaneous causal relations”, which a standard VAR model cannot capture. In this sense, an alternative model that captures both lagged and instantaneous causal relations is the structural VAR (SVAR), a slightly modified model that unifies both SEM and VAR. SEM, VAR and SVAR models are widely used to study linear dependencies among the graph connected time-series. 

\subsection{Nonlinear function approximation}

The main advantages of linear modeling are its simplicity, the low variance of the estimators (at the cost of a higher bias compared to more expressive methods), and the fact that linear estimation problems often lead naturally to convex optimization problems, which can be solved efficiently.

However, there are several challenges related to inferring linear, stationary models from real-world data. Many instances such as financial data, brain signals, industrial sensors, etc. exhibit highly nonlinear interactions, and only nonlinear models have the expressive capacity to capture complex dependencies (assuming that those are identifiable and enough data are provided for the learning). Some existing methods have tried to capture nonlinear interactions using kernel-based function approximators (see  \cite{shen2019nonlinear,money2021online} and references therein).
%\textcolor{red}{Add references to Rohan's paper and 1 or 2 more references there-in}.
In the most general non-linear case, each data variable $y_n[t]$ can be represented as a non-linear function of several multi-variate data time series as:
    \begin{align} \label{eq:nonvar}
	y_n[t] =   h_n(y_{t-1},\ldots, y_{t-P}) + u_n[t],
	\end{align}
    where $y_{t-p} = [y_1[t-p],y_2[t-p],.....,y_N[t-p]]^{T}$, and $h(\cdot)$ is a non-linear function.

However, from a practical perspective, this model is too general to be useful in real applications, because the class of possible nonlinear functions is unrestricted and, therefore, the estimators will suffer from high variance. Notice also that learning such a model would require in general an amount of data that may not be available in realistic scenarios, and requiring a prohibitive complexity. A typical solution is to restrict the the modeling to a subset of nonlinear functions, either in a parametric (NN) or nonparametric (kernel) way.

Our goal in this paper is to learn nonlinear dependencies with some underlying structure making it possible to learn them with limited complexity, with an expressive slightly higher than linear models.

\section{Modelling}
\label{sec:modelling}

The linear coefficients in (1) are tailored to assessing only linear mediating dependencies. To overcome this
limitation, this work considers a non-linear model by introducing a set of node dependent nonlinear functions $\{f_i\}_{i=1}^{N}$. 
Previous works on nonlinear topology identification \cite{shen2019nonlinear,tank2017interpretable,money2021online} estimate nonlinear multivariate models without necessarily assuming linear dependencies in an underlying space; rather, they directly estimate non-linear functions from and into the real measurement space without assuming an underlying structure. In our work, we assume that the multivariate data can be explained as the nonlinear output of a set of observation functions $\{f_i\}_{i=1}^{N}$ with a VAR process as an input. Each function $f_i$ represents a different non-linear distortion at the $i$-th node.

Given data time series, the task is to jointly learn the non-linearities together with a VAR topology in a feature space which is linear in nature, where the outputs of the functions $\{f_i\}_{i=1}^{N}$ belong to. Such functions are required to be invertible, so that sensor measurements can be mapped into the latent feature space, where the linear topology (coefficients) can be used to generate predictions, which can be taken back to the real space through $\{f_i\}_{i=1}^{N}$. In our model, prediction involves the composition of several functions, which can be modeled as neural networks. The nonlinear observation function at each node can be parameterized by a NN that is in turn a universal function approximator \cite{cybenko1989approximation}. 
%\textcolor{red}{ADD REFERENCE TO UNIVERSAL REPRESENTATION THEOREM}. 
Consequently, the topology and non-linear per-node transformations can be seen in aggregation as a DNN, and its parameters can be estimated using appropriate deep learning techniques.

\begin{figure}[h]
\vspace{-0.6cm}
\centering
\includegraphics[width=0.8\textwidth]{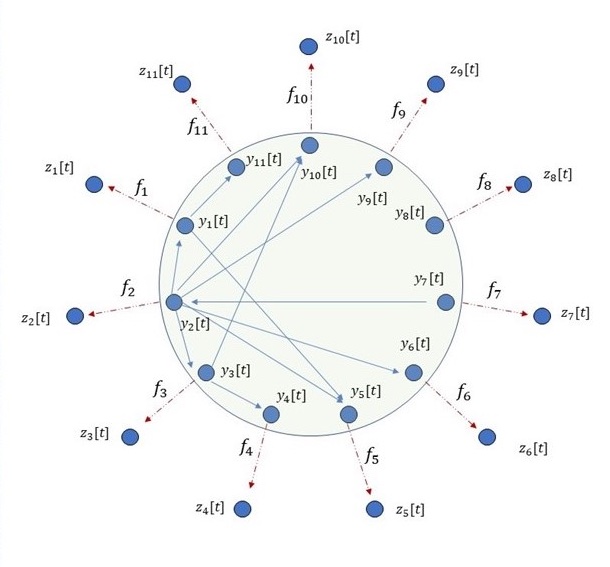}
\vspace{-0.5cm}
\caption{{Causal dependencies among a set of time series are linear in the latent space represented by the green circle. However, the variables in the latent space are not available, only nonlinear observations (output of the functions $f_i$) are available.}\vspace{-0.3cm}} 
\label{fig:swn3}
\end{figure}

The idea is illustrated in Figure \ref{fig:swn3}. The green circle represents the underlying latent vector space. The exterior of the circle is the space where the sensor measurements lie, which need not be a vector space. The blue lines show the linear dependency between the time series inside the latent space. The red line from each time series shows the transformation to the measurement space. Each sensor is associated with a different nonlinear function. Specifically, if $y_i[t]$ denotes the $i$-th time series in the latent space, the measurement (observation) is modeled as $z_{i}[t]=f_{i}\left(y_{i}[t]\right)$. The function $f_i$  is parameterized as a neural network layer with $M$ units, expressed as follows:
\begin{align} 
\label{eq:f_model}
    f_{i}\left(y_{i}\right)=\sum_{j=1}^{M} \alpha_{ij} \sigma\left(w_{ij}y_{i}-k_{ij}\right)+b_{i}
\end{align}

For the function $f_i$ to be monotonically increasing (which guarantees invertibility), it suffices to ensure that $\alpha_{ij}$ and  $w_{ji}$ are positive $\forall j$. The pre-image of $f_i$ is the whole set of real numbers, but the image is an interval $(\ubar{z}_i, \bar{z}_i)$, which is in accordance ro the fact that sensor data are usually restricted to a dynamic range. If the range is not available a priori but sufficient data is available, bounds for the operation interval can be easily inferred.

Let us remark three important advantages in the proposed model: 
\begin{itemize}
    \item It is substantially more expressive than the linear model, while capturing non-linear dependencies with lower complexity than other non-linear models. 
    \item It allows to predict with longer time horizons ahead within the linear latent space. Under a generic non-linear model, the variance of a long-term prediction explodes with the time horizon.
    \item Each non-linear nodal mapping can also adapt and capture any possible drift or irregularity in the sensor measurement, thus, it can directly incorporate imperfections in the sensor measurement itself due to, e.g. lack of calibration.    
\end{itemize}

\subsection{Prediction}

\label{sec:prediction}

Given accurate estimates of the nonlinear functions $\{f_i\}_{i=1}^N$, their inverses, and the parameters of the VAR model, future measurements can be easily predicted. Numerical evaluation of the inverse of $f_i$ as defined in \eqref{eq:f_model} can easily be done with a bisection algorithm. 

Let us define $g_i = f_i^{-1}$. Then, the prediction consists of three steps, the first one being mapping the previous samples back into the latent vector space:
\begin{subequations}
    \begin{equation} \label{eq:forward_g}
        \tilde{y}_{i}[t-p] =g_{i}\left(z_{i}[t-p]\right)
    \end{equation}
Then, the VAR model parameters are used to predict the signal value at time t (also in the latent space):
    \begin{equation} \label{eq:forward_A}
        \hat{y}_{i}[t] =\sum_{p=1}^{p}\sum_{j=1}^{n}  a_{i j}^{(p)} \tilde{y}_{j} [t-p] 
    \end{equation}
Finally, the predicted measurement at each node is obtained by applying $f_i$ to the latent prediction:
    \begin{equation} \label{eq:forward_f}
        \hat{z}_{i}[t] =f_{i}\left(\hat{y}_{i}[t]\right) 
    \end{equation}
\end{subequations}

\begin{figure}[t]
\vspace{-1.7cm}
\hspace{-1.2cm}
\vspace{-0cm}
\includegraphics[width=1.2\textwidth]{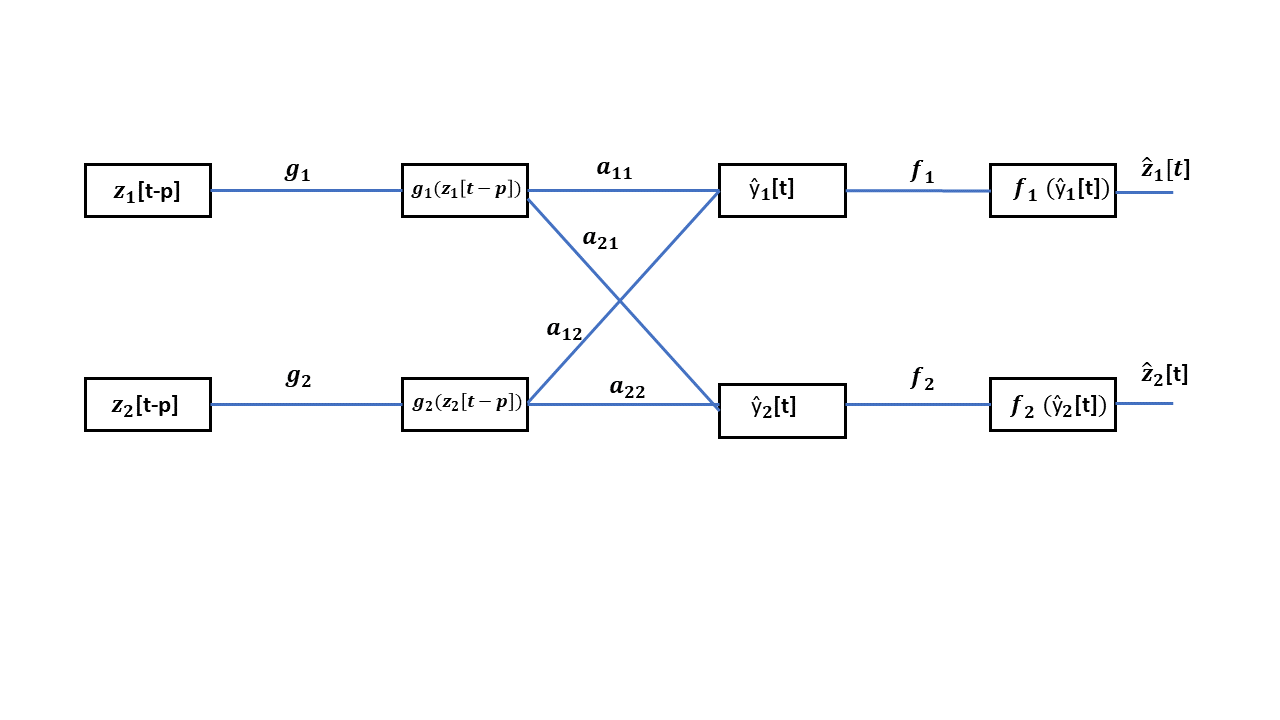}
\vspace{-2.8cm}
\caption{{ Schematic for modeling Granger causality for a toy example with 2 sensors. }} 
\label{fig:example_2sensors}
\end{figure}

These prediction steps can be intuitively visualized as a neural network. The next section formulates an optimization problem intended to learn the parameters of such a neural network. For a simple example with 2 sensors, the network structure is shown in Figure \ref{fig:example_2sensors}.

\section{Problem formulation}

The functional optimization problem consists in minimizing $\|{z}[t]-\hat z\|_{2}^{2}$ (where $z[t]$ is a vector collecting the measurements for all N sensors at time $t$), subject to the constraint of $f_i$ being invertible $\forall i$, and the image of $f_i$ being $(\ubar{z}_i, \bar{z}_i)$. The saturating values can be obtained from the nominal range of the corresponding sensors, or can be inferred from data. 
 
 Incorporating equation (1),  the optimization problem can be written as:
\begin{subequations}
\label{eq:optimization_problem}
\begin{align}
    \min _{f, A} \;\;& \| {z}[t]-f\Big(\sum_{p=1}^{p} A^{(p)}\big[g(z[t-p])\big]\Big) \|_{2}^{2} 
    \\
    \textrm{s. to:}\;\  
    & \sum_{j=1}^{M}\alpha_{j i} 
      = \bar{z}_i %max(z_{i}[t-p])
        - \ubar{z}_i %min(z_{i}[t-p]) 
        \; \forall i
        \label{eq:constraint_alpharange}
    \\
    & b_{i} = \bar{z}_i%min(z_{i}[t-p]) 
        \; \forall i
        \label{eq:constraint_barz}
    \\
    & \alpha_{j i} \geq0 
        \; \forall i, j
        \label{eq:constraint_alphapos}
        \\
    & w_{j i} \geq0 
        ; \forall i, j
        \label{eq:constraint_wpos}
\end{align}
 \end{subequations}
% \textcolor{red}{comment about the max and min saying that the function f saturates at z bar and  below if you know a priori you impose  directly or else infer from the data}     
         
The functional optimization over $f_i$ is tantamount to optimizing over $\alpha_{j i}$,$w_{j i}$,  $k_{j i}$ and $b_{i}$. The main challenge to solve this problem is that there is no closed form for the inverse function $g_i$. This is addressed in the ensuing section.

\iffalse
\begin{lstlisting}
// Algorithm for function g implementation
def g(x,i):
  niter = 10000
  vy = 0
  for j in range(niter):
    vy = vy - (ginverse(vy,i)-x)/gprime(vy,i)  
  return vy
def gprime(x,i): 
  a=0.01
  for j in range(m):
            a = a + alpha[j][i] * sigmoid(x-k[j][i]) * (1-sigmoid(x-k[j][i])) 
  return a
def ginverse(x,i): 
  a = 0   
  for j in range(m):
      a = a + alpha[j][i] * sigmoid(x-k[j][i]) + b[i][0]
  return a
\end{lstlisting}
\fi

%\subsection{Implementation as a neural network}

\section{Learning algorithm}

Without a closed form for $g$, we cannot directly obtaining gradients with automatic differentiation such as Pytorch,%\footnote{One could try to differentiate through all iterations of the Newton method that numerically inverts $f_i$, but there is no guarantee of obtaining a stable gradient estimate, and the computational cost is probably high.}
as is typically done in deep learning with a stochastic gradient-based optimization algorithm. Fortunately, once $\{g_i(\cdot)\}$ is numerically evaluated, the gradient at that point can be calculated with a relatively simple algorithm, derived via implicit differentiation in Sec. \ref{ss:backprop}. Once that gradient is available, the rest of the steps of the backpropagation algorithm are rather standard.

\subsection{Forward equations}
The forward propagation equations are given by the same steps that are used to predict next values of the time series $z$:

\begin{subequations}
    \begin{equation} \label{eq:forward_g}
        \tilde{y}_{i}[t-p] =g_{i}\left(z_{i}[t-p], \theta_{i}\right)
    \end{equation}
    \begin{equation} \label{eq:forward_A}
        \hat{y}_{i}[t] =\sum_{p=1}^{p}\sum_{j=1}^{n}  a_{i j}^{(p)} \tilde{y}_{j} [t-p] 
    \end{equation}
    \begin{equation} \label{eq:forward_f}
        \hat{z}_{i}[t] =f_{i}\left(\hat{y}_{i}[t], \theta_{i}\right) 
    \end{equation}
    \begin{equation} \label{eq:forward_cost}
        C[t] =\sum_{n=1}^{N}\left(z_{n}[t]-\hat{z}_{n}[t]\right)^{2}_\cdot
    \end{equation}
\end{subequations}
Here, the dependency of the nonlinear functions with the neural network parameters is made explicit, where $$\theta_{i}=\left[\begin{array}{l}\alpha_{ i} \\  w_i\\ k_{ i} \\ b_{i}\end{array}\right] \text{ and } 
\alpha_{i}=\left[\begin{array}{c}
\alpha_{i 1} \\
\alpha_{i 2} \\
\vdots \\
\alpha_{i M}
\end{array}\right],w_{i}=\left[\begin{array}{c}
k_{i 1} \\
k_{i 2} \\
\vdots \\
k_{i M}
\end{array}\right], k_{i}=\left[\begin{array}{c}
k_{i 1} \\
k_{i 2} \\
\vdots \\
k_{i M}
\end{array}\right]_. $$

\subsection{Backpropagation equations}
\label{ss:backprop}
The goal of backpropagation is to  calculate the gradient of the cost function with respect to the VAR parameters and the node dependent function parameters $\theta_i$.

%The dimension of the parameter vector $\theta_{i}$ is for the i th sensor  will be (2M+1). 
The gradient of the cost is obtained by applying the chain rule as following:
\begin{equation} \label{chainrule1}
\begin{array}{c}

\frac{d C[t]}{d \theta_{i}}=\sum_{n=1}^{N} \frac{\partial C}{\partial \hat{z}_{n}[t]}  \frac{\hat{z}_{n}[t]}{\partial \theta_{i}} \\
\text { where } \frac{\partial C}{\partial \hat{z}_{n}[t]}=2(\hat{z}_n[t]-z_n[t]) = S_n
\end{array}
\end{equation}

\begin{equation} \label{chainrule2}
\frac{\partial \hat{z}_{n}[t]}{\partial \theta_{i}}=\frac{\partial f_{n}}{\partial \hat{y}_{n}}  \frac{\partial \hat{y}_{n}}{\partial \theta_{i}}+\frac{\partial f_{n}}{\partial \theta_{n}}  \frac{\partial \theta_{n}}{\partial \theta_{i}} \\
\quad 
\end{equation}
$$
\text { where } \frac{\partial \theta_{n}}{\partial \theta_{i}}=\left\{
\begin{array}{l}
I, n = i \\
0, n  \neq i
\end{array}\right.
$$
Substituting equation \eqref{chainrule1} into \eqref{chainrule2} yields
\begin{equation} \label{derivativecost1}
\frac{d C[t]}{d \theta_{i}}=\sum_{n=1}^{N} S_{n}\left(\frac{\partial f_{n}}{\partial \hat{y}_{n}}  \frac{\partial \hat{y}_{n}}{\partial \theta_{i}}+\frac{\partial f_{n}}{\partial \theta_{n}}  \frac{\partial \theta_{n}}{\partial \theta_{i}}\right)_\cdot
\end{equation}
Equation\eqref{derivativecost1} can be simplified as:

\begin{equation} \label{derivativecost2}
    \frac{d C[t]}{d \theta_{i}}
    =
    S_{i}
    \frac{\partial f_{i}}{\partial \theta_{i}}
    +\sum_{n=1}^{N} S_{n}\frac{\partial f_{n}}{\partial \hat{y}_{n}}  \frac{\partial \hat{y_n}}{\partial \theta_{i}}.
\end{equation}

$\text { The next step is to derive } \frac{\partial \hat{y}_{n}}{\partial \theta_{i}} \text { and } \frac{\partial f_{i}}{\partial \theta_{i}}$ of  equation \eqref{derivativecost2}:

\begin{equation} \label{gradientyhat}
    \frac{\partial \hat{y}_{n}[t]}{\partial \theta_i}
    =
    \sum_{p=1}^{P}\sum_{j=1}^{N}  
        a_{n j}^{(p)} 
        \frac{\partial}{\partial \theta_{j}} \tilde{y}_{j}[t-p]
        \frac{\partial \theta_{j}}{\partial \theta_{i}} .
\end{equation}

With $f_{i}^{\prime}\left(z\right)= 
\frac{\partial f_i\left(z, \theta_{i}\right)}{\partial\left(z\right)}, $  
expanding $\tilde{y}_j[t-p]$ in equation \eqref{gradientyhat} changes \eqref{derivativecost2} to:

%\begin{equation} \label{derivativecost3}
%\text { So } \frac{d C[t]}{d \theta_{i}}=S_{i}\left(\frac{\partial f_{i}}{\partial \theta_{i}}\right)+\sum_{p=1}^{P} \sum_{n=1}^{N} S_{n} f_{n}^{\prime}(\hat{y}_n[t])  a_{n i}^{(p)} \frac{\partial}{\partial \theta_{i}} g_{i}\left(z_{i}[t-p],\theta_{i}\right)_\cdot
%\end{equation}

%equation \eqref{derivativecost3} becomes:

\begin{equation} \label{derivativecost4}
 \frac{d C[t]}{d \theta_{i}}=S_{i}\left(\frac{\partial f_{i}}{\partial \theta_{i}}\right)+\sum_{n=1}^{N} S_{n}\left(f_{n}^{\prime}(\hat{y}_{n}[t]) \sum_{p=1}^{P} a_{n i}^{(p)} \frac{\partial}{\partial \theta_{i}} g_{i}\left(z_{i}[t-p],\theta_{i}\right)\right)_\cdot
\end{equation}

Here, the vector 
$$\frac{\partial f_i\left(z, \theta_{i}\right)}{\partial \theta_{i}} %\text{ in equation } \eqref{derivativecost4} 
= \left[
\frac{\partial f_i\left(z, \theta_{i}\right)}{\partial \alpha_{i}} 
\frac{\partial f_i\left(z, \theta_{i}\right)}{\partial w_{i}} 
\frac{\partial f_i\left(z, \theta_{i}\right)}{\partial k_{i}} 
\frac{\partial f_i\left(z, \theta_{i}\right)}{\partial b_{i}} \right]$$ can be obtained by standard or automated differentiation via, e.g., Pytorch \cite{NEURIPS2019_9015}.

However, \eqref{derivativecost4} involves the calculation of $\frac{\partial g_i(z, \theta_{i})}{\partial \theta_{i}}$, which is not straightforward to obtain. Since $g_i(z)$ can be computed numerically, the derivative can be obtained by implicit differentiation, realizing that the composition of $f_i$ and $g_i$ remains invariant, so that its total derivative is zero:

\begin{equation} \label{dfwrttheta1}
\frac{d}{d \theta_{i}}\left[f_i\left(g_i\left(z, \theta_{i}\right), \theta_{i}\right)\right]=0
\end{equation}

\begin{equation} \label{dfwrttheta2}
\Rightarrow \frac{\partial f_i\left(g_i\left(z, \theta_{i}\right), \theta_{i}\right)}{\partial g\left(z, \theta_{i}\right)} \frac{\partial g\left(z, \theta_{i}\right)}{\partial \theta_{i}}+\frac{\partial f_i\left(z, \theta_{i}\right)}{\partial \theta_{i}}=0
\end{equation}

\begin{equation} \label{dfwrttheta3}
\Rightarrow {f^{\prime}_i(g_i(z,\theta_{i}))} \frac{\partial g\left(z, \theta_{i}\right)}{\partial \theta_{i}}+\frac{\partial f_i\left(z, \theta_{i}\right)}{\partial \theta_{i}}=0
\end{equation}

\begin{equation} \label{dgwrttheta1}
\text { Hence } \frac{\partial g_i\left(z, \theta_{i}\right)}{\partial \theta_{i}}=
-\big\{f^{\prime}_i(g_i(z,\theta_{i}))\big\}^{-1}{\left(\frac{\partial f_i\left(z, \theta_{i}\right)}{\partial \theta_{i}}\right)}_\cdot 
\end{equation}
%
%$$\text{where }\frac{\partial g_i\left(z, \theta_{i}\right)}{\partial \theta_{i}} =  \left[\begin{array}{l}
%\frac{\partial g_i\left(z, \theta_{i}\right)}{\partial \alpha_{i}} \\
%\frac{\left.\partial g_i\left(z\right, \theta_{i}\right)}{\partial k_{i}} \\
%\frac{\partial g_i(z,\theta_{i})}{\partial b_{i}}
%\end{array}\right]_\cdot$$
%

The gradient of $C_T$ w.r.t. the VAR coefficient $a^{(p)}_{ij}$ is calculated as follows: 

\begin{equation} \label{dCwrtthetaa}
\frac{d C[t]}{d a^{(p)}_{i j}}=\sum_{n=1}^{N} S_{n}  \frac{\partial f_{n}}{\partial \hat y_{n}}  \frac{\partial \hat{y}_{n}}{\partial a_{i j}^{(p)}}
\end{equation}

\begin{equation} \label{dCwrtthetaa1} \nonumber
\frac{\partial \hat{y}_{n}[t]}{\partial a_{i j}^{(p)}}=\frac{\partial}{\partial a_{i j}^{(p)}} \sum_{p^\prime=1}^{P} \sum_{q=1}^{N} a_{n q}^{(p^\prime)} \tilde{y}_{q}[t-p]
\end{equation}

\begin{equation} \label{dCwrtthetaa2}
\begin{array}{c}
\text { where } 
\frac{\partial a_{n q}^{(p^\prime)}}{\partial a_{i j}^{(p)}}=\left\{\begin{array}{l}
1, n=i, p = p^\prime, \text { and } q=j \\
0, \text {otherwise}
\end{array}\right.
\end{array}
\end{equation}

%\begin{equation} \label{dCwrtthetaa3}
%\frac{\partial \hat{y}_{n}[t]}{\partial a_{i j}^{(p)}}=\left\{\begin{array}{c}
%\sum_{p=1}^{P} \tilde{y}_{j}[t-p], i=n \\
%0, \quad i \neq n
%\end{array}\right.
%\end{equation}

\begin{equation} \label{dCwrtthetaa4}
\frac{d C[t]}{d a_{i j}^{(p)}}=S_i f_{i}^{\prime}\left(\hat{y}_{i}[t]\right) \tilde{y}_{j}[t-p]_\cdot 
\end{equation}

Even though the backpropagation cannot be done in a fully automated way, it can be realized by implementing equations \eqref{dgwrttheta1} and \eqref{derivativecost4} after automatically obtaining the necessary expressions.

\subsection{Parameter optimization}

The elements in $\{A^{(p)}\}_{p=1}^P,$ and $\{\theta_i\}_{i=1}^{N}$ can be seen as the parameters of a NN. Recall from Fig. \ref{fig:example_2sensors} that the prediction procedure resembles a typical feedforward NN as it interleaves component-wise nonlinearities with multidimensional linear mappings. The only difference is that one of the layers computes the inverse of a given function, and its backward step has been derived. Moreover, the cost function in \eqref{eq:optimization_problem} is the mean squared error (MSE).

The aforementioned facts support the strategy of learning the parameters using state-of-the-art NN training techniques. A first implementation has been developed using stochastic gradient descent (SGD) and its adaptive-moment variant Adam \cite{kingma2014adam}. Constraints \eqref{eq:constraint_alpharange}-\eqref{eq:constraint_wpos} are imposed by projecting the output of the optimizer into the feasible set at each iteration.

The approach is flexible enough to be extended with neural training regularization techniques such as dropout %\cite{dropout} 
or adding a penalty based on the L1 or L2 norm of the coefficients, to address the issue of over-fitting and/or promote sparsity. The batch normalization technique can be proposed to improve the training speed and stability. %Last but not least, future developments include improving the interpretability by imposing sparsity. 

\iffalse
\begin{lstlisting}
// Algorithm for back propagation implementation
def backward_propagation(H,alpha,b,A,k,X_train, z_pred):
for i in range(n):
          for p in range(n):
              a1=0
              for j in range(len(X_train)):
                  a1  = a1-2*(X_train[j][p] - z_pred[j][p])*(g(X_train[j][i],i))*(f(H[j][p],p))
              dA[i][p] =a1
    for i in range(n): 
        for p in range(n):                        
            for j in range(len(X_train)):
                dalpha[j][i] = dalpha[j][i] -2*(X_train[j][i] - z_pred[j][i])  *(f(H[j][p],p)*A[i][p]*dalphag(X_train[j][i],i)) -2*(X_train[j][i] - z_pred[j][i])*sigmoid(H[j][i]-k[j][i])
    for i in range(n): 
        for p in range(n):                       
            for j in range(len(X_train)):
                dk[j][i] = dk[j][i] -2*(X_train[j][i] - z_pred[j][i]) * (f(H[j][p],p)*A[i][p]*dkg(X_train[j][i],i)) -2*(X_train[j][i] - z_pred[j][i])*alpha[j][i]*sigmoid(H[j][i]-k[j][i])
                *(1-sigmoid(H[j][i]-k[j][i]))
    for i in range(n): 
        a1 = 0
        for p in range(n):                        
            for j in range(len(X_train)):
                a1  = a1 -2*(X_train[j][i] - z_pred[j][i])  *(f(H[j][p],p)*A[i][p]*dbg(X_train[j][i],i)) -2*(X_train[j][i] - z_pred[j][i])*alpha[j][i]*sigmoid(H[j][i]-k[j][i])
        a1 = db[i][0]            
    return 0

\end{lstlisting}
\fi
\section{Experiments}

The experiments described in this section, intended to validate the proposed method, can be reproduced with the Python code which is available in GitHub at \url{https://github.com/uia-wisenet/NonlinearVAR}

A set of $N=10$ sensors is simulated, and an underlying VAR process of order $P = 2$. The VAR parameter matrices are generated by drawing each weight i.i.d from a standard Gaussian distribution. Matrices $\{A^{(p)}\}_{p=0}^P$ are  scaled down afterwards by a constant that ensures that the VAR process is stable \cite{lutkepohl2005}.

\begin{figure}[h]
\vspace{-0.5cm}
\centering
\includegraphics[width=\textwidth]{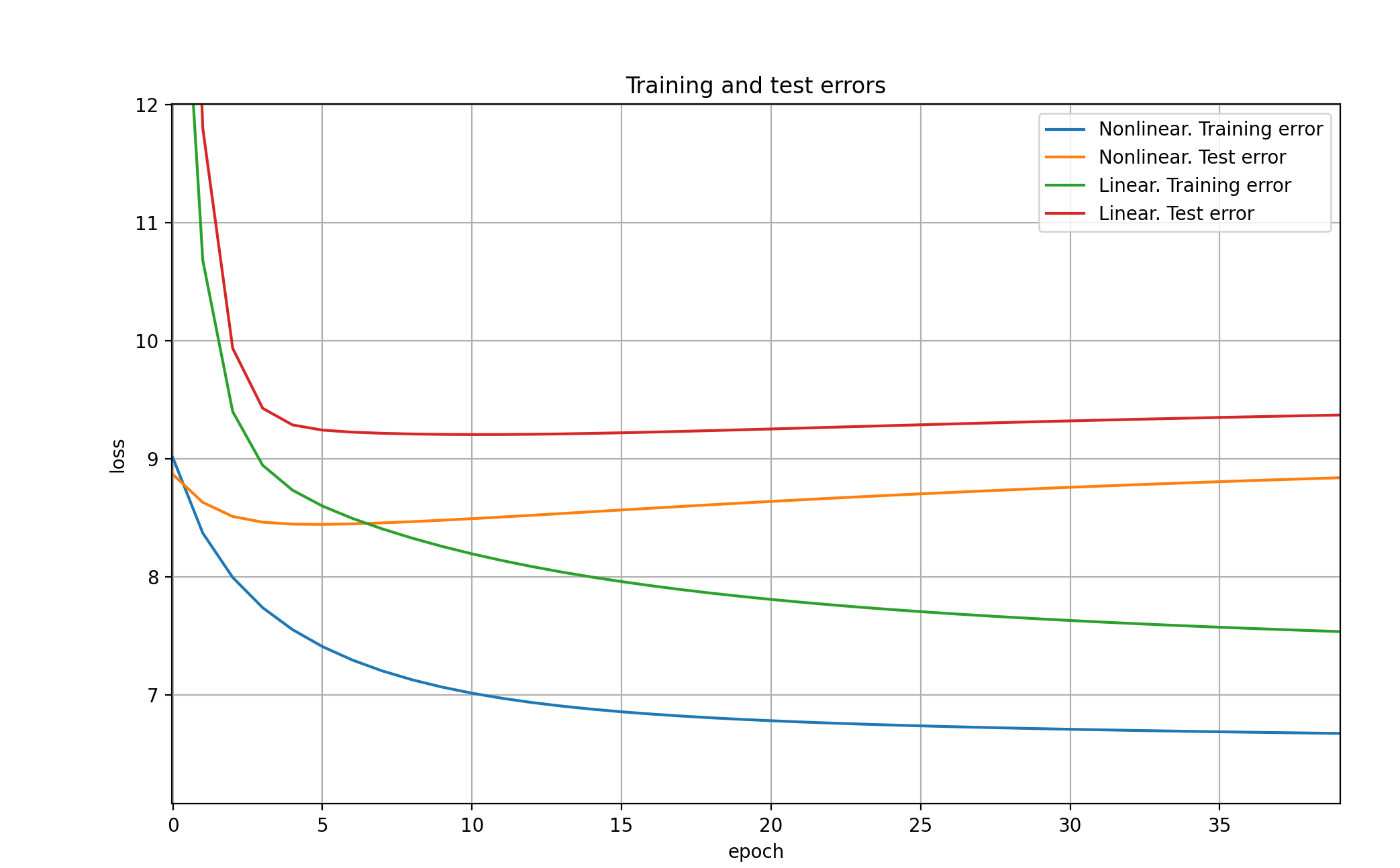}
\caption{Comparison of the proposed method (M=5, P=3) vs. a linear VAR model.% (in blue) 
}
\label{fig:nl_vs_linear4}
\end{figure}

The underlying process samples $\{y[t]\}_{t=1}^T$, where T = 1000, are generated as a realization of the aforementioned VAR process, and the simulated sensor observed values $\{z[t]\}_{t=1}^T$ are obtained as the output of nonlinear observation functions that are also randomly generated. %TODO: explain how

% The values of all parameters involved in the experiments are listed in the captions and legends of the figures.

%In the experiments throughout this section, unless otherwise stated, \revvv{in each MC iteration the following random variables are generated:
%\begin{myitemize}
	%\myitem i) a binary matrix (adjacency of a random graph), 
	%\myitem ii) a set of VAR parameters having the generated matrix as support, and	
	%\myitem iii) a realization of multivariate signals governed by the VAR parameters.
	%\myitem For each signal realization, the proposed algorithms are run and their output is assessed via the metrics defined above, and expectations in (\ref{eq:nmsd}--\,\ref{eq:NMSE}) are taken with respect to realizations of the
	%graph, VAR parameters, and innovation process $\bm u[t]$.
%\end{myitemize}

The proposed nonlinear VAR estimator is analyzed in a stationary setting, and compared to the VAR estimator of the same order. The training and test curves are shown in Fig. \ref{fig:nl_vs_linear4}. It can be observed that, despite the overfitting, the proposed nonlinear model can explain the time series data with significantly lower error.

\section{Conclusion}

A method for inferring nonlinear VAR models has been proposed and validated. The modeling assumption that the observed data are the outputs of nodal nonlinearities applied to the individual time series of a linear VAR process lying in an unknown latent vector space. Since the number of parameters that determine the topology does not increase, the model interpretability remains the same as that with linear VAR modeling, making the proposed model amenable for Granger causality testing and network topology identification. The optimization method, similar to that of DNN training, can be extended with state-of-the-art tools to accelerate training and avoid undesired effects such as convergence to unstable points and overfitting.

\subsubsection{Acknowledgement:}

The authors would like to thank Emilio Ruiz Moreno for helping us manage a more elegant derivation of the gradient of $g_i(\cdot)$.

\bibliographystyle{IEEEtran}
\bibliography{intap}

\begin{thebibliography}{10}
\providecommand{\url}[1]{#1}
\csname url@samestyle\endcsname
\providecommand{\newblock}{\relax}
\providecommand{\bibinfo}[2]{#2}
\providecommand{\BIBentrySTDinterwordspacing}{\spaceskip=0pt\relax}
\providecommand{\BIBentryALTinterwordstretchfactor}{4}
\providecommand{\BIBentryALTinterwordspacing}{\spaceskip=\fontdimen2\font plus
\BIBentryALTinterwordstretchfactor\fontdimen3\font minus
  \fontdimen4\font\relax}
\providecommand{\BIBforeignlanguage}[2]{{%
\expandafter\ifx\csname l@#1\endcsname\relax
\typeout{** WARNING: IEEEtran.bst: No hyphenation pattern has been}%
\typeout{** loaded for the language `#1'. Using the pattern for}%
\typeout{** the default language instead.}%
\else
\language=\csname l@#1\endcsname
\fi
#2}}
\providecommand{\BIBdecl}{\relax}
\BIBdecl

\bibitem{giannakis2018topology}
G.~B. {Giannakis}, Y.~{Shen}, and G.~V. {Karanikolas}, ``Topology
  identification and learning over graphs: Accounting for nonlinearities and
  dynamics,'' \emph{Proceedings of the IEEE}, vol. 106, no.~5, pp. 787--807,
  2018.

\bibitem{chen2018dynamic}
Z.~Chen and S.~V. Sarma, ``Dynamic neuroscience,'' \emph{Chen and SV Sarma,
  Eds. Cham: Springer International Publishing AG}, 2018.

\bibitem{fujita2010granger}
A.~Fujita, P.~Severino, J.~R. Sato, and S.~Miyano, ``Granger causality in
  systems biology: Modeling gene networks in time series microarray data using
  vector autoregressive models,'' in \emph{Advances in Bioinformatics and
  Computational Biology}, C.~E. Ferreira, S.~Miyano, and P.~F. Stadler,
  Eds.\hskip 1em plus 0.5em minus 0.4em\relax Berlin, Heidelberg: Springer
  Berlin Heidelberg, 2010, pp. 13--24.

\bibitem{shen2019nonlinear}
Y.~{Shen}, G.~B. {Giannakis}, and B.~{Baingana}, ``Nonlinear structural vector
  autoregressive models with application to directed brain networks,''
  \emph{IEEE Transactions on Signal Processing}, vol.~67, no.~20, pp.
  5325--5339, 2019.

\bibitem{tank2017interpretable}
A.~Tank, I.~Cover, N.~J. Foti, A.~Shojaie, and E.~B. Fox, ``An interpretable
  and sparse neural network model for nonlinear granger causality discovery,''
  \emph{arXiv preprint arXiv:1711.08160}, 2017.

\bibitem{zaman2020online}
B.~Zaman, L.~M.~L. Ramos, D.~Romero, and B.~Beferull-Lozano, ``Online topology
  identification from vector autoregressive time series,'' \emph{IEEE
  Transactions on Signal Processing}, 2020.

\bibitem{nassif2021automatic}
F.~Nassif and S.~Beheshti, ``Automatic order selection in autoregressive
  modeling with application in eeg sleep-stage classification,'' in
  \emph{ICASSP 2021 - 2021 IEEE International Conference on Acoustics, Speech
  and Signal Processing (ICASSP)}, 2021, pp. 5135--5139.

\bibitem{zhou2021parameter}
R.~Zhou, J.~Liu, S.~Kumar, and D.~P. Palomar, ``Parameter estimation for
  student’s t var model with missing data,'' in \emph{ICASSP 2021 - 2021 IEEE
  International Conference on Acoustics, Speech and Signal Processing
  (ICASSP)}, 2021, pp. 5145--5149.

\bibitem{ioannidis2019semiblind}
V.~N. Ioannidis, Y.~Shen, and G.~B. Giannakis, ``Semi-blind inference of
  topologies and dynamical processes over dynamic graphs,'' \emph{IEEE
  Transactions on Signal Processing}, vol.~67, no.~9, pp. 2263--2274, 2019.

\bibitem{shen2018online}
Y.~Shen and G.~B. Giannakis, ``Online identification of directional graph
  topologies capturing dynamic and nonlinear dependencies,'' in \emph{2018 IEEE
  Data Science Workshop (DSW)}, 2018, pp. 195--199.

\bibitem{money2021online}
R.~Money, J.~Krishnan, and B.~Beferull-Lozano, ``Online non-linear topology
  identification from graph-connected time series,'' \emph{arXiv preprint
  arXiv:2104.00030}, 2021.

\bibitem{farnoosh2017semiparametric}
R.~Farnoosh, M.~Hajebi, and S.~J. Mortazavi, ``A semiparametric estimation for
  the nonlinear vector autoregressive time series model.'' \emph{Applications
  \& Applied Mathematics}, vol.~12, no.~1, 2017.

\bibitem{tank2021neural}
A.~Tank, I.~Covert, N.~Foti, A.~Shojaie, and E.~B. Fox, ``Neural granger
  causality,'' \emph{IEEE Transactions on Pattern Analysis \& Machine
  Intelligence}, no.~01, pp. 1--1, mar 2021.

\bibitem{morioka2021independent}
H.~Morioka, H.~H{\"a}lv{\"a}, and A.~Hyvarinen, ``Independent innovation
  analysis for nonlinear vector autoregressive process,'' in
  \emph{International Conference on Artificial Intelligence and
  Statistics}.\hskip 1em plus 0.5em minus 0.4em\relax PMLR, 2021, pp.
  1549--1557.

\bibitem{jin2020correlated}
M.~{Jin}, M.~{Li}, Y.~{Zheng}, and L.~{Chi}, ``Searching correlated patterns
  from graph streams,'' \emph{IEEE Access}, vol.~8, pp. 106\,690--106\,704,
  2020.

\bibitem{yanuar2014estimation}
F.~Yanuar, ``The estimation process in bayesian structural equation modeling
  approach,'' \emph{Journal of Physics: Conference Series}, vol. 495, p.
  012047, 04 2014.

\bibitem{granger1988concept}
W.~Granger~Clive, ``Some recent developments in a concept of causality [j],''
  \emph{Journal of Econometrics}, 1988.

\bibitem{lutkepohl2005}
H.~Lütkepohl, \emph{New Introduction to Multiple Time Series Analysis}.\hskip
  1em plus 0.5em minus 0.4em\relax Springer, 2005.

\bibitem{cybenko1989approximation}
G.~Cybenko, ``Approximation by superpositions of a sigmoidal function,''
  \emph{Mathematics of control, signals and systems}, vol.~2, no.~4, pp.
  303--314, 1989.

\bibitem{NEURIPS2019_9015}
\BIBentryALTinterwordspacing
A.~Paszke, S.~Gross, F.~Massa, A.~Lerer, J.~Bradbury, G.~Chanan, T.~Killeen,
  Z.~Lin, N.~Gimelshein, L.~Antiga, A.~Desmaison, A.~Kopf, E.~Yang, Z.~DeVito,
  M.~Raison, A.~Tejani, S.~Chilamkurthy, B.~Steiner, L.~Fang, J.~Bai, and
  S.~Chintala, ``Pytorch: An imperative style, high-performance deep learning
  library,'' in \emph{Advances in Neural Information Processing Systems 32},
  H.~Wallach, H.~Larochelle, A.~Beygelzimer, F.~d\textquotesingle
  Alch\'{e}-Buc, E.~Fox, and R.~Garnett, Eds.\hskip 1em plus 0.5em minus
  0.4em\relax Curran Associates, Inc., 2019, pp. 8024--8035. [Online].
  Available:
  \url{http://papers.neurips.cc/paper/9015-pytorch-an-imperative-style-high-performance-deep-learning-library.pdf}
\BIBentrySTDinterwordspacing

\bibitem{kingma2014adam}
D.~P. Kingma and J.~Ba, ``Adam: A method for stochastic optimization,''
  \emph{arXiv preprint arXiv:1412.6980}, 2014.

\end{thebibliography}

\end{document}